# Local mechanical properties of electrospun fibers correlate to their internal nanostructure

*Andrea Camposeo,[†,||,‡,*] Israel Greenfeld,[⊥,‡,*] Francesco Tantussi,[#] Stefano Pagliara,[†,§] Maria Moffa,[†,||] Francesco Fuso,[#,∇] Maria Allegrini,[#,∇] Eyal Zussman,[⊥] Dario Pisignano,[†,||,¶]*

[†]National Nanotechnology Laboratory of Istituto Nanoscienze-CNR, via Arnesano, I-73100 Lecce (Italy)

[||]Center for Biomolecular Nanotechnologies @UNILE, Istituto Italiano di Tecnologia, via Barsanti, I-73010 Arnesano, LE (Italy)

[⊥]Department of Mechanical Engineering, Technion - Israel Institute of Technology, Haifa 32000, Israel

[#] Dipartimento di Fisica "Enrico Fermi" and CNISM, Università di Pisa, Largo Bruno Pontecorvo 3, I-56127 Pisa (Italy)

[∇]Istituto Nazionale di Ottica INO-CNR, Sezione di Pisa, Largo Bruno Pontecorvo 3, I-56127 Pisa (Italy)

[¶]Dipartimento di Matematica e Fisica "Ennio De Giorgi", Università del Salento, via Arnesano I-73100 Lecce, (Italy)

KEYWORDS. NANOFIBERS, NEAR-FIELD MICROSCOPY, YOUNG'S MODULUS, CONJUGATED POLYMERS.

ABSTRACT. The properties of polymeric nanofibers can be tailored and enhanced by properly managing the structure of the polymeric molecules at the nanoscale. Although electrospun polymer fibers are increasingly exploited in many technological applications, their internal nanostructure, determining their improved physical properties, is still poorly investigated and understood. Here, we unravel the internal structure of electrospun functional nanofibers made by prototype conjugated polymers. The unique features of near field optical measurements are exploited to investigate the nanoscale spatial variation of the polymer density, evidencing the presence of a dense internal core embedded in a less dense polymeric shell. Interestingly, nanoscale mapping the fiber Young's modulus demonstrates that the dense core is stiffer than the polymeric, less dense shell. These findings are rationalized by developing a theoretical modeling and simulations of the polymer molecular structural evolution during the electrospinning process. This model predicts that the stretching of the polymer network induces a contraction towards the jet center of the network with a local increase of the polymer density, as observed in the solid structure. The found complex internal structure opens interesting perspective for improving and tailoring the molecular morphology and multifunctional electronic and optical properties of polymer fibers.

Fiber-shaped materials are the building blocks of many natural systems[1,2] and the enabling components of some of the most important modern technologies.[3-6] The advent of nanotechnologies has enabled the synthesis of micro- and nano-scale fibers by a variety of approaches, with a prominent control on shape and composition.[7] Experimental and theoretical research efforts have evidenced enhanced electronic, optical and mechanical properties of these innovative, almost 1-dimensional (1D) nanomaterials compared to the bulk counterpart.[7-9] Among 1D nanomaterials, polymer nanofibers deserve particular attention, because the use of polymers is continuously increasing in many fields, especially in low-end applications, where cost considerations prevail over performances. In this framework, polymeric 1D nanomaterials offer both low costs and physical properties enhanced by the nanoscopic morphology and peculiar assembly of macromolecules within nanofibers.[10-12] In particular, by reducing the fiber diameter below a critical value, an increase of the Young's modulus can be obtained,[10] demonstrating the possibility of tailoring the mechanical properties by controlling the geometry and supramolecular assembly in polymer nano-systems. Moreover, the peculiar packing of organic semiconductors in 1D nanostructures allows improved charge mobilities, polarized emission, enhanced amplified spontaneous emission and non-linear optical properties to be observed, and a control of energy transfer phenomena to be obtained.[9-14] Therefore, predicting and managing the resultant polymer supramolecular assembly and the nanofiber internal structure is becoming increasingly relevant, aiming to ultimately optimize the performance of polymer-based systems and devices through smart engineering of the different processing steps.

Polymer nanofibers are mainly fabricated by elongating and stretching a polymer solution or melt by mechanical, capillary or electrostatic forces.[11, 15, 16] This may result in extended chain conformation, very different with respect to standard solution or melt processing methods (spincoating, casting, rapid prototyping etc.). The access to the fiber internal nanostructure of the polymer macromolecules is however challenging. So far, studies on the inner features of 1D polymeric systems have utilized small-area electron diffraction (SAED),[17] transmission electron microscopy,[18] infrared[19] and Raman[20,21] spectroscopies, however either the limited spatial resolution or the inability to probe molecular orientation have prevented to resolve the internal structure of the analyzed systems, having sub-micron characteristic features.

In this work, we investigate the complex internal structure of conjugated polymer nanofiber materials. In particular, the nanoscale spatial variation of the fiber Young's modulus, and of the polymer density determined by near-field measurements, evidences the presence of a stiff and dense internal core with typical size of nearly 30% of the fiber diameter, embedded in a softer and less dense polymeric shell. These findings are supported by theoretical modeling and simulations of the molecular structural evolution during the elongational flow of semidilute polymer solutions at the base of electrospinning, which predict substantial stretching of the polymer network, accompanied by its contraction towards the jet center, as observed in the solid structure. The understanding and prediction of the internal structure of active fiber materials can be very important for the design and realization of novel advanced functional materials.

To our aim, a prototype conjugated polymer is used that constitutes an unequalled tool for probing optically the fiber internal nanostructure with nm-resolution. Fibers are made by electrospinning the poly[2-methoxy-5-(2-ethylhexyl-oxy)-1,4-phenylene-vinylene] (MEH-PPV), which is largely used in lasers,[22] field effect transistors[23] and light emitting diodes.[24] Randomly- and uniaxially-oriented free-standing, flexible mats of fibers (Fig. 1a) are produced by dissolving the polymer in a mixture of good and poor solvents (see Methods).[25] The fibers emit visible light peaked at 605 nm as shown in Fig. 1b, where we also show the temperature dependence of the emission. The photoluminescence (PL) peak blue-shifts by about 10 nm upon increasing temperature, which can be attributed to a decreased conjugation length due to excitation of torsional and liberation modes.[26, 27] More importantly, the blue-shift observed in fibers is smaller than that in thin films by about a factor two,[26] evidencing reduced sensitivity to torsional distortions. This suggests irregular molecular assembly in the fibers compared to the film, which motivates to investigate their internal nanoscale structure much more in depth. Indeed, the stretching process, whose dynamics is determined by competing forces related to the applied electric field and molecular interactions (surface tension and viscoelesticity), as well as by rapid solvent evaporation, can result in complex internal nanostructuring.[28]

To study the effects of such phenomena on individual fibers, we determine their mechanical and densitometric properties by nanoscale indentation experiments and scanning near-field optical microscopy (SNOM). The local Young's modulus of a fiber deposited on quartz can be obtained with atomic force microscopy (AFM) by measuring the nanoscale deformation induced by a controlled load, applied along a direction perpendicular to the fiber longitudinal axis and to the substrate (Fig. S3 in the Supporting Information). The mechanical response of the nanofiber upon indentation depends on its elastic properties, which are mainly related to the local density, degree of crystallinity and arrangement of the polymer molecules. Interestingly, the MEH-PPV fibers feature a spatially non-uniform effective elastic modulus ($E_{fiber}$), whose resulting value is affected by the polymer structure underlying the indentation region. Overall, in an axial region (whose width is roughly 30% of the fiber diameter), $E_{fiber}$ is about twice the value measured in the peripheral region which constitutes the external layer of each fiber. However, due to the low thickness of the fibers (typically <200 nm), these measurements are affected by the mechanical properties of the substrate underneath.[29,30] To rule out such effects, indentation experiments are better performed on the cross-sectional surface of cleaved fibers. To this aim, we firstly embed MEH-PPV fibers in a photo-curable polymer, and freeze the resulting solid composite in liquid nitrogen. Following careful fracturing, the fiber cross-sectional surfaces are clearly visible both by emission confocal microscopy and by AFM (Fig. 2a,b). Examples of Young's modulus maps measured on the fiber cross sections are shown in Fig. 2c-e, where data clearly evidence the presence of a stiffer internal region nearby the fiber longitudinal axis, extending over about 30% of the cross sectional area. This axial region exhibits a Young's modulus up to 80-120 MPa, larger than that in the surrounding sheath by about a factor 2.

This has to be clearly correlated to the internal nanostructure and density, which we also investigate by SNOM in order to probe simultaneously morphology and optical properties with sub-wavelength resolution.[31-34] Figure 3a displays the map of the transmittance, $T(x, y) = I_s(x, y)/I_{sub}$, obtained by raster scanning the sample, measuring the intensity of the transmitted light, $I_s(x, y)$, and normalizing to the light transmitted by the transparent regions of a quartz substrate ($I_{sub}$). The measured $T$ values are superimposed to the simultaneously acquired fiber topography, and used to calculate the average absorption coefficient along the local beam path. In Figure 3b we display the absorption coefficient, $\alpha / \alpha_{max}$, normalized the maximum absorption value measured in the single fiber ($\alpha_{max}$ in the range 3-3.5×10$^4$ cm$^{-1}$). It is remarkable that the map showing the spatial variation of the absorption coefficient is *not* flat, as would be in case of homogeneous distribution of the absorbers. Instead, comparison of the line profiles of the absorption coefficient and the fiber height (Fig. 3c) clearly indicates a higher concentration of absorbing chromophores at the fiber core. Such a non-uniform distribution of the absorbing chromophores has been observed in all the investigated fibers (see Supporting Information). Overall, both mechanical and optical data evidence that the electrospun conjugated polymer fibers are characterized by a core-sheath structure with a denser and stiffer core, which can significantly impact on technological and optoelectronic applications.

In order to rationalize the origin of such a complex internal structure of the nanomaterial, we develop a model of the polymer elongational dynamics during electrospinning, where the flow of the solution jet exerts strong stretching forces. Owing to inherent bonding defects, which substitute rigid conjugated links by flexible tetrahedral links along the chain backbone, the conjugated macromolecules can be described as flexible chains[35] with specific adjustments pertaining to their high segmental aspect ratio.[36] The conjugated polymer chain is so treated as a linear, flexible, freely-jointed chain, whose rigid segments are chain sections between neighboring bonding defects. Scaling is used to incorporate the interactions relevant to the solvent type and to describe the entangled polymer network conformation in the semidilute solution.[35] An example of a simulated polymer network at rest is shown in Fig. 4a. During electrospinning, each subchain is acted upon by the hydrodynamic force induced by the solvent, as well as by the entropic forces applied by its neighboring subchains. The resulting conformational evolution has been previously modeled for fully flexible chains, using a beads-and-springs lattice model and a 3D random walk simulation.[35] This is readily applicable to conjugated subchains, using as input the calculated initial network mesh size ($\xi_0$=20 nm) and number of segments per subchain ($N_s$=14), corresponding to the polymer volume fraction ($\varphi$=0.025), and assuming a defects concentration of 10% of monomers together with the jet velocity. Since evaporation is negligible at the early stage of electrospinning, the jet velocity can be derived (see details in the following) from the measured radius, $a$, of the jet (Fig. 4b). The simulation provides the dependence of the simulated polymer network radius, $a_p$, on the longitudinal spatial coordinate, $z$ (Fig. 4b). The polymer subchains contract laterally as a consequence of the redistribution of probabilities between the axial and radial directions of the random walk. The lateral contraction of individual subchains affects the conformation of the whole polymer network, narrowing its radius $a_p$ faster than the narrowing of the jet radius $a$. The simulated conformation of the whole network and its evolution along the jet (Figure 4b) demonstrate the dominant effect of axial stretching and lateral contraction, while only a negligible effect of radial hydrodynamic compression. The network compacts around the jet center, thereby increasing polymer concentration near the center. Experimental evidence of this effect has been reported for optically-inert polymers such as poly(ethylene oxide) and poly(methylmethacrylate), by measuring the polymer jet absorption profile during electrospinning with fast X-ray phase contrast imaging,[37] whereas it was previously unexplored in active, light-emitting or conductive nanofibers. In fact, our model generalizes the stretching and compacting phenomenon[35,37] for all types of linear polymers, using the degree of chain flexibility as a controlling parameter. In the case of semi-flexible conjugated polymers whose backbone structure is rigid, the bonding defects concentration determines chain flexibility, whereas fully flexible polymer chains are a particular case of the model with null defects concentration. Consequently, the model and simulation predict that the stretching phenomenon should be prominent in conjugated polymers because of their longer rigid segments. Hence, MEH-PPV is an excellent choice as system allowing us to measure for the first time the nanostructure of solid fibers by optical means, owing to its high absorption as well as expectedly higher traces of the effects of electrospinning-induced stretching.

The network conformation during electrospinning depends on the balance between stretching and evaporation.[37] Dominant evaporation can cause rapid solidification of the jet surface, retarding evaporation from the core and resulting in a tubular structure.[28, 38] On the other hand, dominant strain rates will cause higher polymer density in the center due to stretching. Our model shows that the stretching of conjugated chains occurs earlier than in fully flexible chains, and one can therefore expect a dense core in the solid fiber. Indeed, the distance from the needle where full chain extension is accomplished is below 1 mm for semi-flexible conjugated polymers, as demonstrated by overlaying the simulated polymer network on the image of the actual jet (Fig. 4b). Moreover, the theoretical modeling of the network shows that the jet radius reduction ratio, at the position where full extension is reached, is lower by a factor of typically 2-5 (depending on the solvent quality) compared to fully flexible chains, confirming an earlier network stretching in conjugated polymers.

In addition, crystallization is enhanced in regions of strong stretching and alignment. Interchain interaction and $\pi$-$\pi$ stacking are known to lead to high extent of local crystallinity.[39] When neighboring chain sections are aligned in the same direction, they correlate to each other according to Onsager's rods theory, and may eventually crystallize. This phenomenon will be more pronounced in conjugated polymers with longer rigid chain sections between bonding defects. The model shows that, unlike flexible polymers, conjugated chains intermix within a single correlation volume in the network, increasing the probability of interchain overlap. The model specifically predicts that such correlation is likely to occur during electrospinning of MEH-PPV with typical production-induced bonding defects concentration (5-10% of monomers), at the solution concentrations used in our experiments.

Here, the measurements of the material properties of as-spun MEH-PPV solid nanofibers provide convincing evidence that the polymer matrix conformation described for the liquid phase of the jet is essentially retained in the solid nanofiber. In particular, SNOM measurements (Fig. 3) show higher optical absorption at the fiber center and lower absorption closer to its boundary, indicative of higher polymer concentration at the fiber core. The regions of lower concentration close to the boundary have a large fraction of free volume and are likely porous, possibly even encouraging nucleation and growth of crystalline structures.[40]

Moreover, traces of an early solidification of a skin during the spinning[41] can be seen in the slight absorption rise very close to the fiber boundary and on its surface (visible for instance in Fig. 3b). These observations suggest that during electrospinning the solvent content at the jet core is low as a result of network stretching and inward contraction, whereas closer to the jet boundary the solvent content is high and evaporation through the solidified skin leaves voids and porosity in the inner matrix close to the boundary. This is consistent with the measured spatial variation of the Young's modulus (Fig. 2), since lower values are measured far from the fiber axis, where a less dense polymer network is present with higher free-volume content. This process-induced core-sheath structure impacts on many physical properties, determining, for instance, an increase of the effective conjugation length in conjugated polymer nanofibers.[42]

## CONCLUSIONS

In summary, the elongational dynamics of polymer semidilute solutions under electrostatic fields is predicted to include a fast axial stretching of the polymer network accompanied by a radial contraction toward the core, resulting in a higher polymer concentration and axial orientation at the fiber center. Our modeling shows that this morphology should be more pronounced in the semi-flexible conjugated polymers due to their longer rigid chain segments, but evidence from X-ray imaging of electrospinning jets indicates that it is also expected in fully flexible polymers. As demonstrated by the SNOM analysis and the AFM indentation measurements, the polymer conformation during the electrospinning process is retained in the solid matrix. This process-induced core-sheath structure impacts on many physical properties, determining, for instance, an increase of the Young's modulus close to the fiber core. In perspective, the found graded-density internal structure and the different mechanical properties of the core and the sheath of polymer fibers open interesting opportunities for many applications. In organic semiconductors, the presence of a core with close-packed molecules can improve charge transport, whereas the sheath with less dense molecules can determine the suitable conditions to enhance amplification of the light guided in the fiber. Both charge transport and light amplification can be therefore improved in a single nanostructure. For scaffold applications, the complex internal structure can be exploited to engineering multifunctional fibers, where the high density core can provide enhanced mechanical strength and/or feed stimuli (electrical, thermal, etc.), whereas the porous external layer can be a suitable soft substrate for cell adhesion, contaminant removal, or drug delivery.

**METHODS**

**Nanofiber production**. The nanofibers are produced by electrospinning a solution of MEH-PPV (MW 380,000 g/mol, American Dye Source Inc., Baie-d'Urfé, Canada), dissolved in dimethyl sulfoxide and tetrahydrofuran (1:4 w:w, see Supplementary Information). A 70 μM polymer solution is stored into a 1.0 mL plastic syringe tipped with a 27-gauge stainless steel needle, and injected at the end of the needle at a constant rate of 10 μL/min by a microprocessor dual drive syringe pump (33 Dual Syringe Pump, Harvard Apparatus Inc., Holliston, MA). The positive lead from a high-voltage supplier (XRM30P, Gamma High Voltage Research Inc., Ormond Beach, FL) is connected to the metal needle applying a bias of 5 kV. The collector is made of two Al stripes biased at a negative voltage of –6 kV and positioned at a mutual distance of 2 cm and at a distance of 6 cm from the positively charged needle. All the electrospinning experiments are performed at room temperature with air humidity in the range 40-50%. Aligned arrays of free-standing fibers are deposited across metallic stripes and then collected on a 1×1 cm$^2$ quartz substrate. Arrays of uniaxially aligned MEH-PPV nanofibers are also fabricated by using a rotating collector.

**Polymer jet imaging**. For imaging the polymer jet profile, a stereomicroscope (Leica MZ 12.5) and a high speed camera (Photron, FASTCAM APX RS, 1024 pixel ×1024 pixel, 10000 frame s$^{-1}$) are used. A typical collected single frame image is shown in Fig. 4b. The dependence of the jet velocity, $v$, and radius, $a$, on the axial coordinate z, is given by the following relation:[35,37]

$$\frac{v}{v_0} \cong \left(\frac{a}{a_0}\right)^{-2} \cong \left(1+\frac{z}{z_0}\right)^{2\beta}, \tag{1}$$

Given the initial velocity $v_0$ = 5.8 mm/s and radius $a_0$ = 96 μm, fit of the jet radius data yields $z_0 = 22$ μm, and $\beta = 0.94$.

**Nanofiber Characterization.** Fluorescence confocal microscopy is performed by using the A1R MP confocal system (Nikon), coupled to an inverted microscope (Eclipse Ti, Nikon). The fibers are excited by an Ar+ ion laser ($\lambda_{exc}$=488 nm) through an oil immersion objective with numerical aperture of 1.4. PL spectra are collected by exciting the MEH-PPV fibers with a diode laser ($\lambda_{exc}$=405 nm) and collecting the emission by an optical fiber coupled to a monochromator, equipped with a Charge Coupled Device detector (YobinYvon). The fiber samples are mounted in a He closed-cycle cryostat under vacuum ($10^{-4}$ mbar) for variable temperature measurements.

AFM and mechanical compression experiments. AFM imaging is performed by using a Multimode system equipped with a Nanoscope IIIa electronic controller (Veeco Instruments). The nanofiber topography is measured in Tapping mode, utilizing Si cantilevers featuring a resonance frequency of 250 kHz. To map the local Young's modulus of the nanofibers (see Supporting Information for details), force-distance curves are collected by using non conductive, Au-coated Silicon Nitride cantilevers with a nominal spring constant of 0.32 N/m, tip radius of 20 nm and resonant frequency of 52.5 kHz. The fiber is supported underneath by the substrate, assuring no bending or buckling during measurements, and the force is applied perpendicularly to the fiber longitudinal axis and to the substrate. The system is calibrated by measuring the force-distance curve of a stiff sample (Si/SiO$_2$, quartz). For mapping the local mechanical properties on the fiber cross-sectional surface, arrays of uniaxially aligned MEH-PPV fibers are embedded in a photo-curable polymer (NOA68, Norland Products Inc.), that is cured by exposure to UV light for 3 minutes. The curing UV intensity is kept at about 1 mW/cm$^2$ to avoid degradation of the active polymer. The samples are frozen in liquid nitrogen and fractured along a direction perpendicular to the fiber alignment axis. Samples are then inspected by confocal and AFM microscopies, in order to select those showing smooth cross-sectional surfaces for subsequent mechanical measurements.

**SNOM analysis**. Properties at nanoscale are investigated with a scanning near-field optical microscope. The instrument operates in the emission-mode: the sample interacts with the near-field produced by a tapered optical fiber probe (Nanonics) featuring a 50 nm diameter apical aperture (nominal). The system allows the fiber topography (i.e. height profile) to be measured simultaneously to optical transmission in each scan, by the shear-force method. This allows a topography map [$h(x,y)$] to be obtained, which is then used as a local measurement of the fiber thickness. A semiconductor laser with wavelength $\lambda$=473 nm, coupled to the tapered fiber, is used to measure the local absorption of the nanofibers. To this aim, the signal transmitted by the sample is collected by an aspheric lens and sent onto a miniaturized photomultiplier (Hamamatsu R-5600), connected to a lock-in amplifier. In order to avoid any artifact related to variations of the fiber thickness, the absorption coefficient is calculated as: $\alpha(x, y) = - ln[T(x, y)]/h(x, y) = \sigma \rho(x, y)$, where $h(x, y)$ indicates the local nanofiber thickness, deducible from the topography map measured simultaneously to the optical transmission map (Fig. 3a and S2, see Supporting Information for more details). Linear absorption is assumed as dominant and the Lambert-Beer law is used to estimate absorption, in turn related to the absorption cross section at the incident laser wavelength ($\sigma$) and to the local density of absorbing chromophores, $\rho(x,y)$.

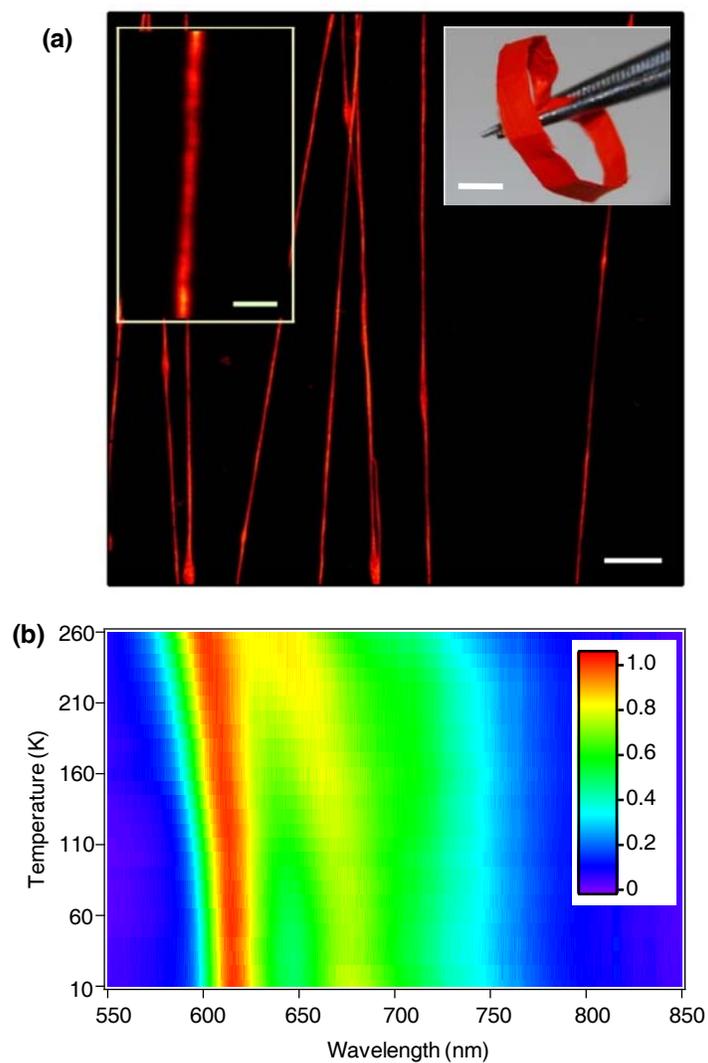

**Figure 1.(a)** Fluorescence confocal micrographs of MEH-PPV fibers. Scalebar: 10 μm. Left inset scalebar: 2 μm. Right inset: photograph of a uniaxially-oriented nanofiber mat (scalebar: 4 mm). **(b)** MEH-PPV fiber emission spectra *vs* sample temperature. Color scale: normalized PL intensity.

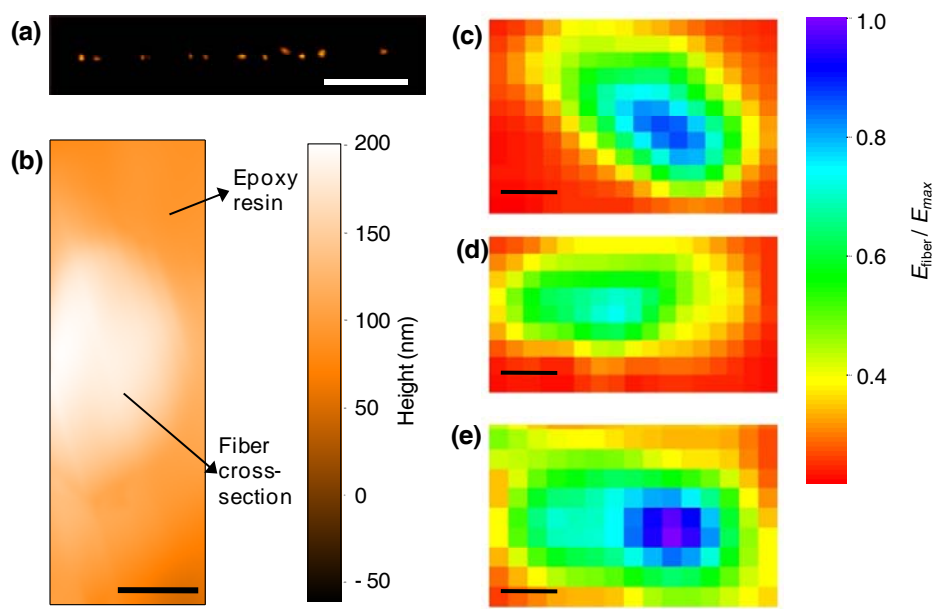

**Figure 2.(a)** Confocal emission image of the cross-sectional surface of an array of MEH-PPV fibers. Scalebar: 10 μm. **(b)** Tapping mode topography micrograph of the cross-sectional surface of an individual MEH-PPV nanofiber (bright region). The fiber is embedded in a UV-cured polymer. Scalebar: 200 nm. **(c)-(e)** Examples of cross-sectional Young's modulus (normalized to the maximum value, $E_{max}$) maps measured by AFM indentation measurements (force-distance curves). The Orange-red regions correspond to the cured polymer embedding the fibers. Scalebars: (c) 500 nm, (d) 200 nm and (e) 300 nm, respectively.

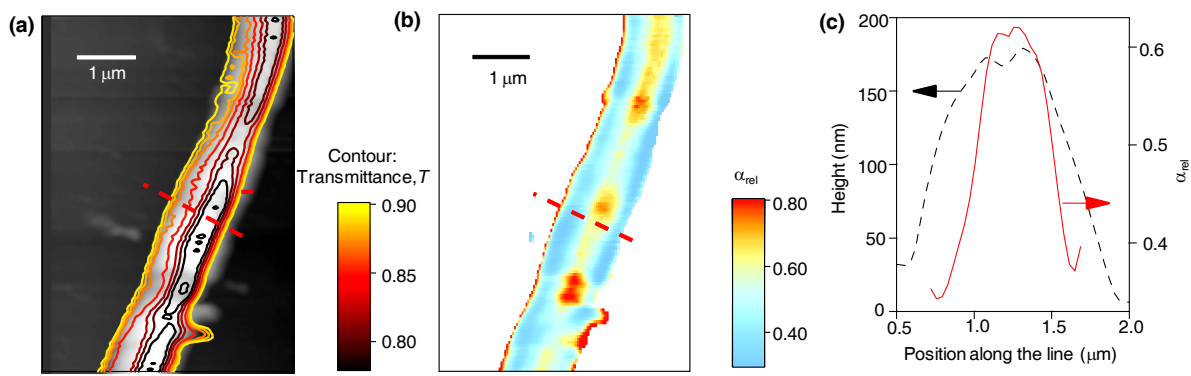

**Figure 3.(a)** Contour plot of the SNOM transmission superimposed to the corresponding topography. The topography map is produced by the shear-force method during the scan of a single MEH-PPV fiber deposited on quartz, and the optical transmission is acquired simultaneously by collecting the signal passing through the sample. Transmission is averaged over all polarization states of the near-field probing radiation. The color scale refers to the contour plot. **(b)** Map of the nanoscale variation of optical absorption. **(c)**, Line profile analysis showing the cross-sections of the topography (dashed line) and the corresponding relative optical absorption, $\alpha/\alpha_{max}$ (continuous line) along the dashed segment in maps **(a)** and **(b)**.

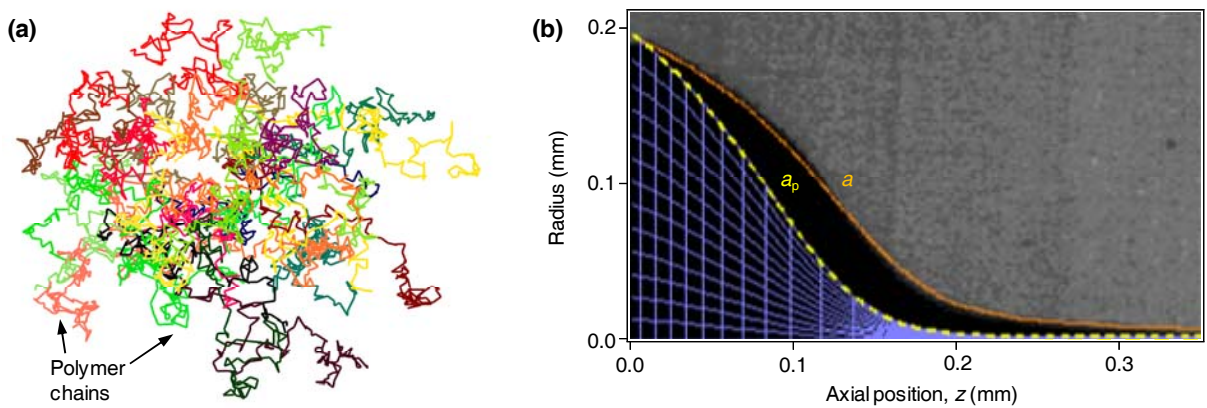

**Figure 4.** (a) Example of a section of a network at rest, made of 30 conjugated polymer chains, each consisting of 146 segments (MW=380,000 g/mol). The size of the network section is about 150 nm, and its average mesh size is 20 nm. (b) Image of the measured steady state jet profile and corresponding jet radius a vs. axial position $z$ (continuous line). The modelled polymer network radius, $a_p$ vs. $z$, is also shown (dashed line), together with the network mesh (viewed mesh density is diluted ×300 in each direction). The maximal jet radius is larger than the needle internal radius, $a_0$=96 μm, due to wetting of the needle face. Electric field = 1.8 kV/cm, flow rate = 10 μL/min, MEH-PPV volume fraction $\phi = 0.025$.

ASSOCIATED CONTENT

**Supporting Information**. A supporting document with additional technical details on nanofiber morphological and mechanical properties, Young's modulus mapping and SNOM analysis is included as a separate PDF file. This material is available free of charge via the Internet at [http://pubs.acs.org](http://pubs.acs.org).

AUTHOR INFORMATION

**Corresponding Authors**

* E-mail: andrea.camposeo@nano.cnr.it, green_is@netvision.net.il, dario.pisignano@unisalento.it

**Present Addresses**

§Current address: Sector of Biological and Soft Systems, Department of Physics, Cavendish Laboratory, University of Cambridge, J.J. Thomson Avenue, Cambridge CB3 0HE, (United Kingdom).

**Author Contributions**

‡ These authors contributed equally to this work.

ACKNOWLEDGMENT

V. Fasano and G. Potente are acknowledged for confocal and SEM images, respectively. The authors also gratefully thank S. Girardo for high-speed imaging of the polymer jet and E. Caldi for assistance in the SNOM measurements. We gratefully acknowledge the financial support of the United States-Israel Binational Science Foundation (BSF Grant 2006061), the RBNI-Russell Berrie Nanotechnology Institute, and the Israel Science Foundation (ISF Grant 770/11). The

Published in Nano Letters, 13:5056-5062, doi: 10.1021/nl4033439 (2013).

research leading to these results has received funding from the European Research Council under the European Union's Seventh Framework Programme (FP/2007-2013)/ERC Grant Agreement n. 306357 (ERC Starting Grant "NANO-JETS").

# SUPPORTING INFORMATION

# Local mechanical properties of electrospun fibers correlate to their internal nanostructure

*Andrea Camposeo,[†,‖,‡,*] Israel Greenfeld,[⊥,‡,*] Francesco Tantussi,[#] Stefano Pagliara,[†,§] Maria Moffa,[†,‖] Francesco Fuso,[#,∇] Maria Allegrini,[#,∇] Eyal Zussman,[⊥] Dario Pisignano,[†,‖,¶]*

[†]National Nanotechnology Laboratory of Istituto Nanoscienze-CNR, via Arnesano, I-73100 Lecce (Italy)

[‖]Center for Biomolecular Nanotechnologies @UNILE, Istituto Italiano di Tecnologia, via Barsanti, I-73010 Arnesano, LE (Italy)

[⊥]Department of Mechanical Engineering, Technion - Israel Institute of Technology, Haifa 32000, Israel

[#] Dipartimento di Fisica "Enrico Fermi" and CNISM, Università di Pisa, Largo Bruno Pontecorvo 3, I-56127 Pisa (Italy)

[∇] Istituto Nazionale di Ottica INO-CNR, Sezione di Pisa, Largo Bruno Pontecorvo 3, I-56127 Pisa (Italy)

[¶] Dipartimento di Matematica e Fisica "Ennio De Giorgi", Università del Salento, via Arnesano I-73100 Lecce, (Italy)

1. **Nanofiber Morphology**

Figure S1 displays the MEH-PPV fiber morphology as obtained by scanning electron microscopy (SEM) and atomic force microscopy (AFM). SEM analysis is performed by using a Nova NanoSEM 450 field emission system (FEI) operating with an acceleration voltage of 5 kV and an aperture size of 30 μm. A thin layer of Cr (<10 nm) is thermally evaporated on top of the samples before SEM imaging. Figures S1a,b display the SEM images of MEH-PPV fibers deposited on quartz substrates and then analysed by near-field optical analysis. The fibers have mean diameters of about 500 nm.

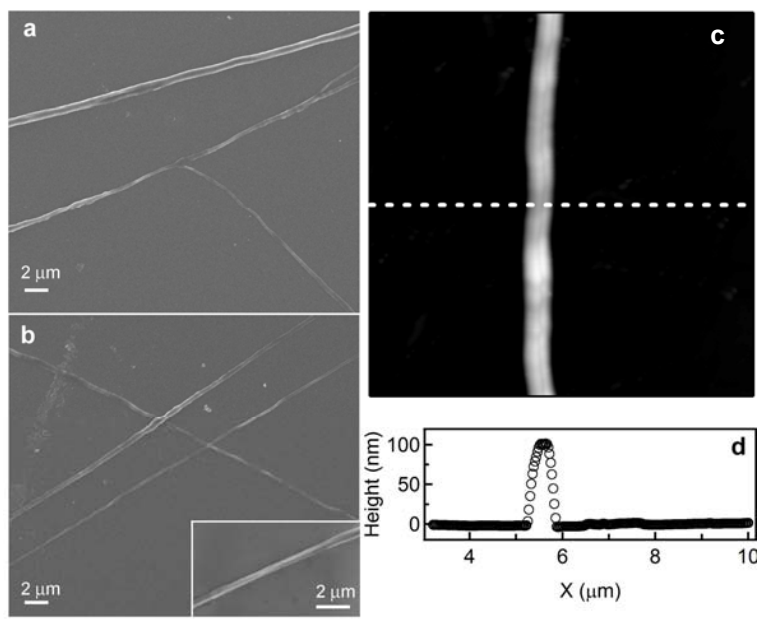

**Figure S1**. (a, b) SEM images of MEH-PPV fibers. (c, d) AFM topographic map (c) and height profile (d) of a MEH-PPV fiber (width 450 nm and height 100 nm.).

The surface topography of the nanofibers is investigated by AFM, employing a Multimode head (Veeco Instruments, Plainview, NY) equipped with a Nanoscope IIIa controller and operating in tapping mode. Phosphorous-doped Si tips are employed, with an 8-10 nm nominal curvature radius and a resonant frequency of 250 kHz. Figures S1c,d display a typical AFM topography map of a MEH-PPV fiber, evidencing a ribbon-shape. Similar results are obtained by measuring the fiber topography by the shear-force method with the scanning near field optical microscope (Fig. S2).

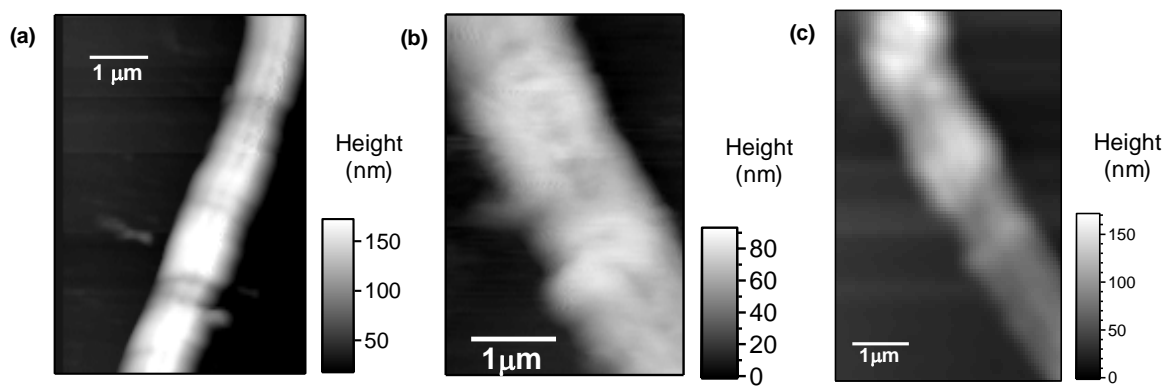

**Figure S2.** Examples of topography maps of MEH-PPV fibers obtained by the shear-force method. The map reported in (a) corresponds to the topography of the fiber displayed in Fig. 3. The fiber height is < 170 nm, whereas the width is about 1 μm, in accordance with the ribbon shape of the fiber evidenced by both AFM and SEM measurements (Fig. S1).

## 2. Force-Indentation measurements.

The nanoscale spatial variation of the nanofiber elastic modulus is measured by acquiring force ($F_{load}$) vs distance curves, by using a Multimode AFM system equipped with a Nanoscope IIIa electronic controller (Veeco Instruments). The force vs. distance curves are then converted in force vs. deformation plots ($F_{load}$ vs. δ).[S1,S2] The dependence of the applied load on the deformation of the sample (δ) is approximated by the Hertz model:[S3]

$$F_{load} = \left(\frac{4}{3}\sqrt{R}\right)\left(\frac{1-\nu_t^2}{E_t} + \frac{1-\nu_{fiber}^2}{E_{fiber}}\right)^{-1} \delta^{\frac{3}{2}} \quad (S1)$$

where $R$ is the tip radius, $\nu_t$ and $\nu_f$ are the Poisson's ratio of the cantilever ($\nu_t = 0.27$) and of the fiber ($\nu_f = 0.35$), respectively, and $E_t$ and $E_{fiber}$ are the Young's modulus of the cantilever ($E_t = 160$ GPa) and of the nanofiber, respectively. The nanofiber Young's modulus is obtained by fitting the force vs. indentation curves to Eq. S1.[S4]

Indentation measurements on the surface of fibers (Fig. S3) are performed by applying the load, $F_{load}$, along a direction perpendicular to the quartz substrate and to the fiber longitudinal axis, assuring the absence of bending or buckling of the fiber during the measurement. Indeed, AFM images of the investigated region, acquired before and after indentation measurements, do not evidence variations of the fiber morphology and position. Due to the finite thickness of the fiber deposited on the quartz substrate, these measurements can be affected by the presence of the stiffer substrate. B. Cappella *et al.*[S5] have reported a dependence of the measured effective elastic modulus on the thickness of the polymer film deposited on glass. In particular they have observed an increase of the effective elastic modulus *upon decreasing* the film thickness, since for thinner films the indentation measurement is sensitive also to the mechanical properties of the substrate.

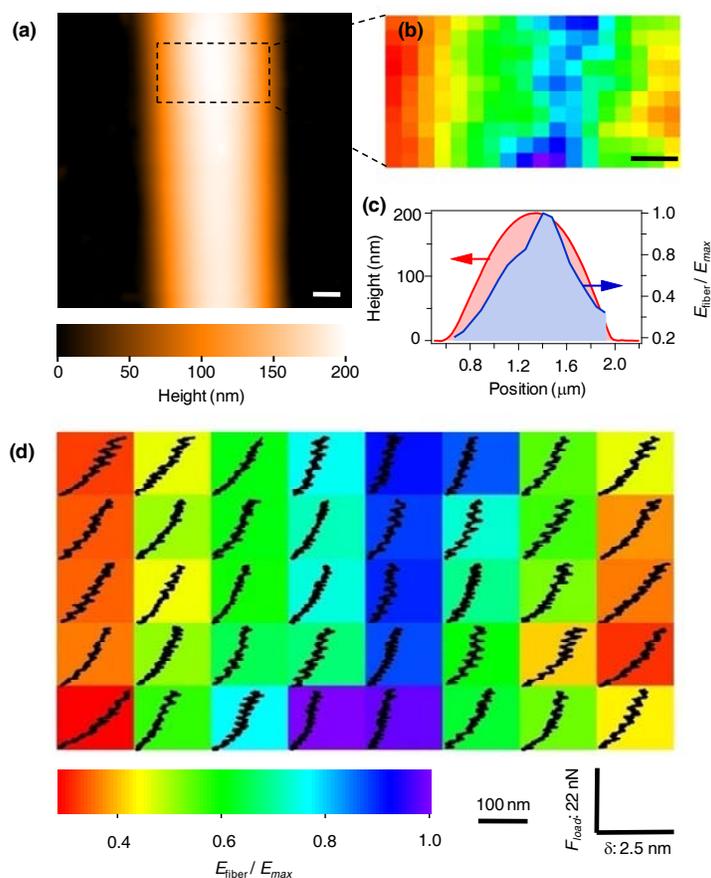

**Figure S3. (a)** AFM micrograph of a single MEH-PPV spun fiber. Scalebar: 250 nm. **(b)** Corresponding Young's modulus ($E_{fiber}$) map, normalized to the maximum value ($E_{max}$). Scalebar: 200 nm, color scale shown in the bottom of the Figure. The map is obtained by determining the force-distance curves in the region highlighted by a dashed box in **(a)**. **(c)** Line profiles showing the cross-sections of the topography (red continuous line) and of $E_{fiber}$ (blue continuous line). **(d)** Example of applied load ($F_{load}$) vs. deformation ($\delta$) curves measured in different points of the fiber surface. Each pixel area is 140×140 nm$^2$, and the pixel color shows the local normalized Young's modulus. The curves shown in each pixel have higher slopes for stiffer regions, according to (x, y) axes ($\delta$ and $F_{load}$), respectively, shown in the bottom-right corner. The overall analyzed area is highlighted in **(a)** by a dashed box.

They also proposed a semi-empirical analysis that allows to obtain the mechanical properties of the polymer film, taking into account its finite thickness.[S6] The force-distance measurements performed on a fiber deposited on quartz substrate provide therefore an effective elastic modulus, possibly affected by the substrate. In particular, for a fiber composed by uniformly distributed polymer, *larger* effective values are expected at the fiber border, due to the reduced thickness and to the relatively major contribution from the substrate. Instead, we find a *decrease* of the elastic modulus at the fiber border (Fig. S3), that can be related to the presence of a softer fiber sheath, as confirmed by the fiber cross-section measurements discussed in the main paper (Fig. 2). In Fig. S4, two examples of force *vs.* indentation ($\delta$) curves, measured at the fiber core and sheath, respectively, are shown.

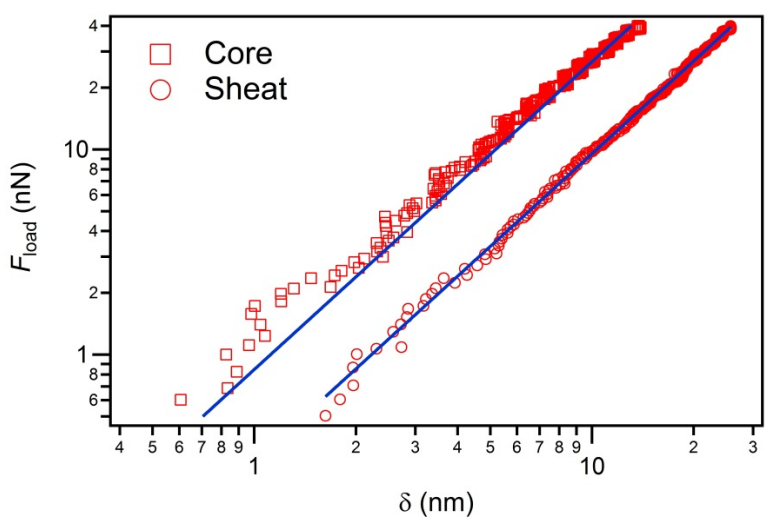

**Figure S4**. Examples of force vs. indentation ($\delta$) curves measured on the fiber cross-section surface in the core (squares) and sheath (circles). The difference of the resulting $E_{fiber}$ is evidenced by the different intercepts of the curves with the $F_{load}$ axis in the bi-logarithmic plot. The continuous lines are fits to the data by Eq. S1.

3. **SNOM measurements**

The spatial variation of the polymer density in the fibers is evaluated by near-field absorption microscopy, a measurement allowing to estimate the absorption coefficient, that depends on the local density of the absorbing chromophores, according to the Lambert-Beer law. In order to obtain maps of the absorption coefficient ($\alpha$), the light transmitted through the fiber illuminated by the optical near field of a tapered fiber is measured simultaneously to its topography. This is accomplished by the shear-force method,[S7] allowing both the fiber-sample distance to be kept constant and the fiber height profile to be obtained in each scan. Examples of fiber topography maps obtained by this method are shown in Fig. S2. The map of the absorption coefficient is then calculated as: $\alpha(x, y) = -\ln[T(x, y)]/h(x, y)$, where $T(x,y)$ is the map of the fiber transmission coefficient and $h(x,y)$ is the local, measured fiber thickness, which is fully taken into account in this way. Examples of transmittance and absorption coefficient maps obtained in various MEH-PPV fibers are shown in Fig. S5.

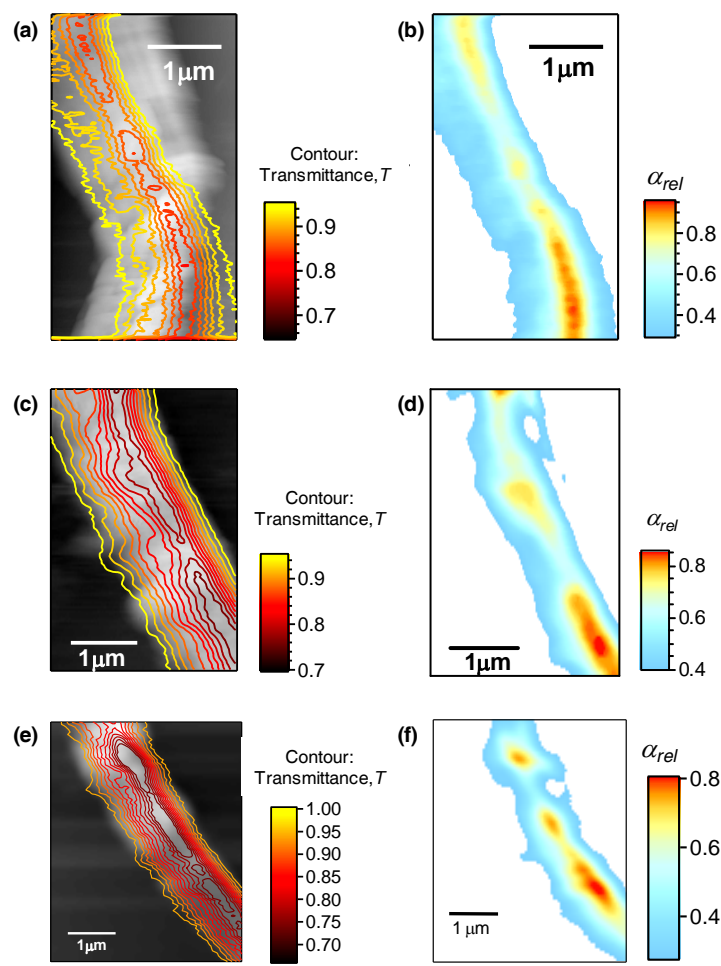

**Figure S5**. (a, c, e) Examples of fiber topography maps, with superimposed contour plots of the SNOM transmission data, and corresponding maps of the nanoscale variation of optical absorption (b, d, f) respectively.

**References.**

S1. Tan, S.; Sherman Jr., R. L.; Ford, W. T. Nanoscale compression of polymer microspheres by atomic force microscopy. *Langmuir* **2004**, *20*, 7015-7020.

S2. Touhami, A.; Nysten, B.; Dufrene, Y. F. Nanoscale mapping of the elasticity of microbial cells by atomic force microscopy. *Langmuir* **2003**, *19*, 4539-4543.